\documentclass[floatfix,superscriptaddress,a4paper,
               showpacs,showkeys,nofootinbib,preprint]{revtex4}
\usepackage{epsfig}
\usepackage{latexsym}
\usepackage{xspace}
\usepackage[T1]{fontenc}
\usepackage[utf8]{inputenc}
\usepackage{indentfirst}
\usepackage{enumerate}
\usepackage{color}
\usepackage{tikz}

\usepackage{amsmath}
\usepackage{amssymb}
\usepackage{mathtools}
\usepackage[english]{babel}
\usepackage{url}
\definecolor{orange}{RGB}{237,169,50}
\definecolor{red}{RGB}{227,76,80}
\definecolor{blue}{RGB}{66,133,187}
\definecolor{green}{RGB}{42,114,38}
\newcommand{\dashedLine}{\tikz[baseline=-0.5ex]\draw [line width=4pt, dashed] (0,0) -- (0.5,0);}

\newcommand{\solidLine}{\tikz[baseline=-0.5ex]\draw [ultra thick] (0,0) -- (0.5,0);}

\newcommand{\eV}{\ensuremath{\mbox{e\kern-0.1em V}}\xspace}
\newcommand{\GeV}{\ensuremath{\mbox{Ge\kern-0.1em V}}\xspace}
\newcommand{\MeV}{\ensuremath{\mbox{Me\kern-0.1em V}}\xspace}
\newcommand{\GeVc}{\ensuremath{\mbox{Ge\kern-0.1em V}\!/\!c}\xspace}
\newcommand{\GeVcc}{\ensuremath{\mbox{Ge\kern-0.1em V}\!/\!c^2}\xspace}
\newcommand{\AGeV}{\ensuremath{A\,\mbox{Ge\kern-0.1em V}}\xspace}
\newcommand{\AGeVc}{\ensuremath{A\,\mbox{Ge\kern-0.1em V}\!/\!c}\xspace}
\newcommand{\MeVc}{\ensuremath{\mbox{Me\kern-0.1em V}/c}\xspace}

\def\avg#1{\langle{#1}\rangle}

\newcommand{\CernVM}{\textsc{Cern\-\kern-0.05emVM}\xspace}

\newcommand{\NASixtyOne}{NA61\slash SHINE\xspace}

\def\avg#1{\langle{#1}\rangle}

\begin{document}

\title{Eliminating \emph{volume} fluctuations in fixed-target heavy-ion experiments}

\author{M. Mackowiak-Pawlowska}
\affiliation{Faculty of Physics, Warsaw University of Technology, Warsaw, Poland}
\email{majam@cern.ch}

\author{M. Naskręt}
\affiliation{University of Wroclaw, Wrocław, Poland}

\author{M. Gazdzicki}
\affiliation{Geothe-University Frankfurt am Main, Germany}
\affiliation{Jan Kochanowski University, Kielce, Poland}

\begin{abstract}
Experimental and theoretical studies of fluctuations in nucleus-nucleus interactions at high energies have started to play a major role in understanding of the concept of strong interactions. The elaborated procedures have been developed to disentangle different processes happening during nucleus-nucleus collisions. The fluctuations caused by a variation of the number of nucleons which participated in a collision are frequently considered the unwanted one. The methods to eliminate these fluctuations in fixed-target experiments are reviewed and tested. 
They can be of key importance in the following ongoing fixed-target heavy-ion experiments:
\NASixtyOne at the CERN SPS, STAR-FT at the BNL RHIC, BM\@N at JINR Nuclotron, HADES at the GSI SIS18 
and in future experiments such as NA60+ at the CERN SPS, CBM at the FAIR SIS100, JHITS at J-PARC-HI MR.

\end{abstract}

\pacs{25.75.q, 25.75.Nq, 24.60.Ky}

\keywords{heavy-ion collisions, fluctuations, fixed-target experiments}
\maketitle


\section{Introduction}
\label{sec:introduction}

Measuring event-by-event fluctuations is the focus of numerous experimental programmes on nucleus-nucleus collisions at high energies.
Nowadays, the leading motivation is the possibility to discover the critical point of strongly interacting matter and a need to understand how the onset of deconfinement influences event-by-event 
fluctuations. The recent reviews can be found in Refs.~\cite{Gazdzicki:2015ska,Bzdak:2019pkr,Gazdzicki:2020jte}.

Fluctuations in high energy collisions are significantly influenced by fluctuations in the amount of matter
(\emph{volume}) and energy involved in a collision, 
as well as global and local conservation laws.
These fluctuations are unwanted effects in the search for the critical point and the study of the onset of deconfinement. 

In this paper methods to remove the influence of the \emph{volume} fluctuations in fixed target experiments are reviewed and tested.
They can be of key importance in the following ongoing fixed-target heavy-ion experiments:
\NASixtyOne~\cite{Abgrall:2014xwa} at the CERN SPS, 
STAR-FT~\cite{Odyniec:2019kfh} at the BNL RHIC, 
BM\@N~\cite{Kekelidze:2018nyo} at the JINR Nuclotron,
HADES~\cite{Agakishiev:2009am} at the GSI SIS18,
and in the future experiments such as 
NA60+~\cite{Agnello:2018evr} at the CERN SPS,
CBM~\cite{Ablyazimov:2017guv} at the FAIR SIS100, 
JHITS~\cite{Sako:2019hzh} at J-PARC-HI MR.

The paper is organized as follows: 
Section~\ref{sec:WNM} introduces the reference model - the Wounded Nucleon Model (WNM)~\cite{Bialas:1976ed} - used here to test the influence of the \emph{volume} fluctuations. This section also introduces extensive, intensive and strongly intensive measures of fluctuations~\cite{Gazdzicki:1992ri,Gorenstein:2011vq} and their 
\emph{volume} dependence within WNM.  
The main features of typical fixed-target and collider experiments with respect to fluctuation measurements and the \emph{volume} fluctuations are summarized in Sec.~\ref{sec:FixedvsCollider}. 
Two methods used to eliminate the effect of the \emph{volume} fluctuations in fixed-target experiments are introduced and compared using  WNM in Sec.~\ref{sec:MethodsTests}.
The summary concludes the paper.

\section{Wounded Nucleon Model, extensive and intensive quantities}
\label{sec:WNM}

\begin{figure}[ht]
	\centering
	\includegraphics[width=0.40\textwidth]{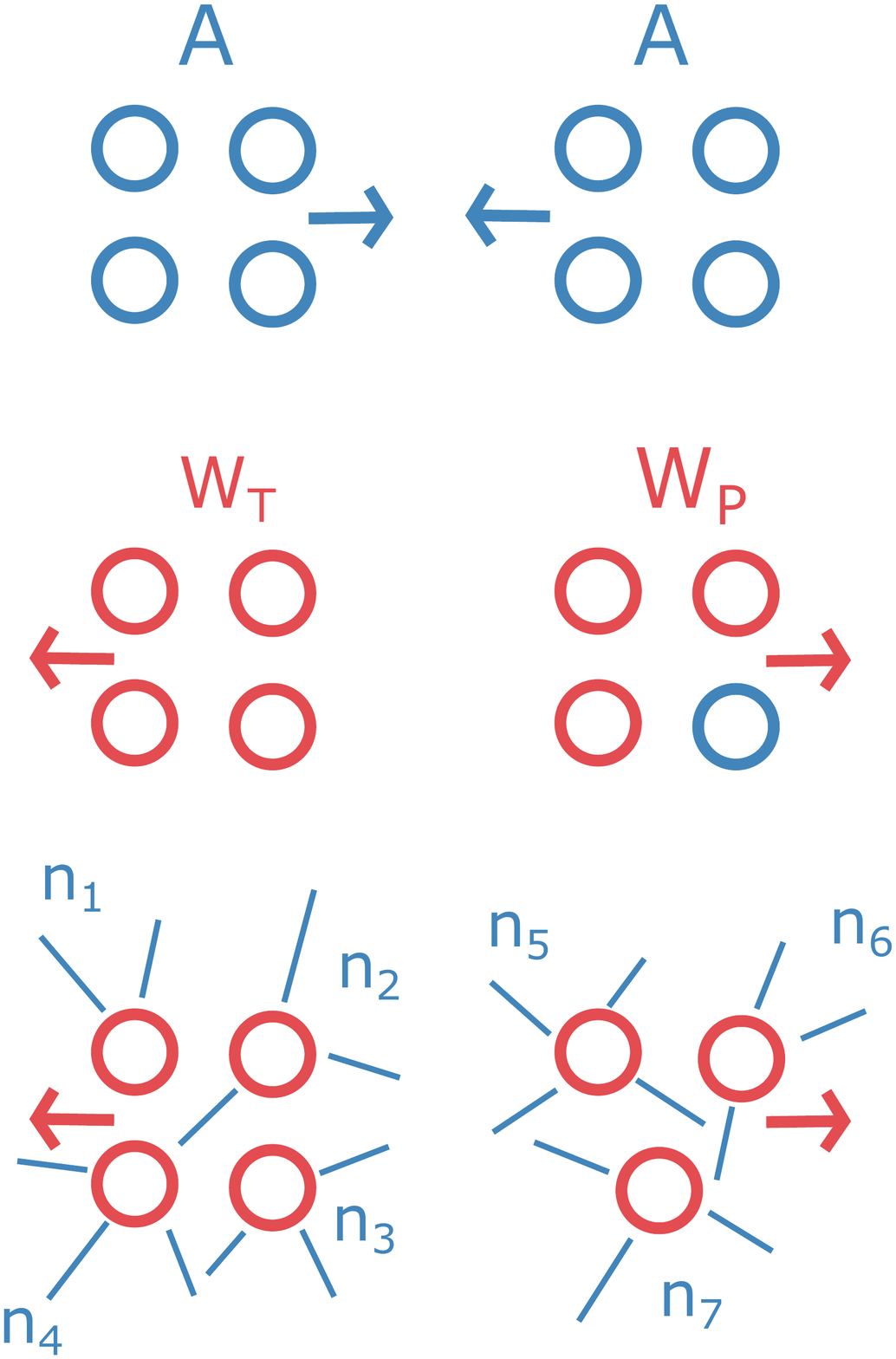}
	\caption[]
	{
		The sketch of particle production process in nucleus-nucleus collisions according to the Wounded Nucleon Model~\cite{Bialas:1976ed}.
		Projectile and target nuclei with nuclear mass number $A_P$ and $A_T$ (here $A = A_P =A_T = 4$)
		collide. $W_T$ (here $ W_T = 4 $) target wounded nucleons and $W_P$ (here $ W_P = 3 $)
		projectile wounded nucleons produce $N$ particles, where $N$ is given by the sum over all wounded nucleons of particle multiplicities $n_i$ from a single wounded nucleon,
		$N = \sum_{i=1}^{7} n_i$.
	}
	\label{fig:wnm}
\end{figure}

Using the the Wounded Nucleon Model~\cite{Bialas:1976ed} is probably the simplest way to introduce fluctuations of the amount of matter involved in a collision
and their impact on the fluctuations of produced particles.
The model was proposed in 1976
as the late child of the S-matrix period~\cite{Gazdzicki:2012sj}.
It assumes that particle production in nucleon-nucleon 
and nucleus-nucleus collisions is an incoherent superposition 
of particle production from wounded nucleons. The wounded nucleons are the ones which interacted inelastically and which number is calculated using straight line trajectories of nucleons.
The properties of wounded nucleons are independent of the size of the colliding
nuclei, e.g., they are the same in \emph{p+p} and Pb+Pb collisions at the same
collision energy per nucleon.
Within WNM, the number of wounded nucleons plays the role of \emph{volume}. 
These assumptions are graphically illustrated in Fig.~\ref{fig:wnm}.

The extensive quantity is proportional to the system \emph{volume}, which
in the WNM is represented by $W$.
Let a random variable $A$ measured for each collision be defined as a sum of corresponding random variables $a_i$ for wounded nucleons:
\begin{equation}
A = a_1 + a_2 \dots + a_W~.
\label{ea:ext}
\end{equation}
For example, $a_i$ can be particle multiplicity produced by $i$-th wounded nucleon $n_i$ and then $A$ is collision multiplicity, $N = \sum_{i=1}^{W} n_i$.

The k-th order moment of the probability distribution of $A$, $P(A)$, is defined as
	\begin{equation}
		\langle A^{k} \rangle = \sum_{A} A^{k}P(A).
    \label{eq:moments}
	\end{equation} 
Then the extensive quantities which correspond to $A$ are cumulants of $A$
by
\begin{eqnarray}
	\kappa_{1}[A]=\langle A\rangle, \\
	\kappa_{2}[A]=\langle \delta A^{2}\rangle= Var[A], \\
	\kappa_{3}[A]=\langle \delta A^{3}\rangle, 	\\
	\kappa_{4}[A]=\langle \delta A^{4}\rangle-3\langle\delta A^{2}\rangle \\ 
    \dots~,\nonumber
\label{eq:cumulants}    
\end{eqnarray}
where $\langle\delta A^{k} \rangle=\langle(A-\langle A \rangle)^{k}\rangle$.
The first and the second cumulants are referred to as the mean and variance of $A$, respectively. The third and fourth cumulants are related to skewness, $S = \kappa_{3} / \kappa_{2}^{3/2}$ and kurtosis, $\kappa = \kappa_{4} / \kappa_{2}^{2}$, respectively.
By definition, cumulants are proportional to $W$.

An intensive quantity is the quantity which is independent of \emph{volume}.
Clearly, the ratio of two extensive quantities is the intensive quantity. 
For example, the ratio of the two first cumulants referred to as scaled variance is an intensive quantity:
		\begin{equation}
			\omega[A] = \kappa_{2}[A] / \kappa_{1}[A].
        \label{eq:omega}
		\end{equation}
Other frequently used intensive quantities which involve third and fourth moments of $A$ are:
\begin{equation}
\  \kappa_{3}[A] / \kappa_{2}[A], \quad  \kappa_{4}[A] / \kappa_{2}[A],
\label{Eq:HM}
\end{equation}
sometimes denoted as $S\sigma$ and $\kappa\sigma^{2}$, respectively.
For any probability distribution $P(W)$,
the scaled variance calculated within the WNM reads~\cite{Gorenstein:2011vq}:
\begin{equation}
\omega[A] = \omega[A]_W + \langle A \rangle / \langle W \rangle \cdot \omega[W] ~, \\ 
\label{eq:wnm:varN}
\end{equation}
where $\omega[N]_W$ stands for the scaled variance at any fixed number of wounded nucleons and
$W = W_P + W_T$. 
The first component of Eq.~\ref{eq:wnm:varN} is considered the wanted one and
it is independent of the \emph{volume} fluctuations. However,  the second component is unwanted and it is proportional to the scaled variance of the $W$ distribution.
Corresponding expressions for higher order moments are given in Ref.~\cite{Begun:2016sop}.

It is worth noting that similar relations are valid within Statistical Models of an Ideal Boltzmann gas within the Grand Canonical Ensemble SM(IB-GCE)~\cite{Gorenstein:2011vq}.
Then, in the equations above, the number of wounded nucleons $W$ should be replaced by the gas volume $V$.

\section{Fixed-target versus collider experiments}
\label{sec:FixedvsCollider}

\begin{figure}[ht]
	\centering
		\includegraphics[width=0.60\textwidth]{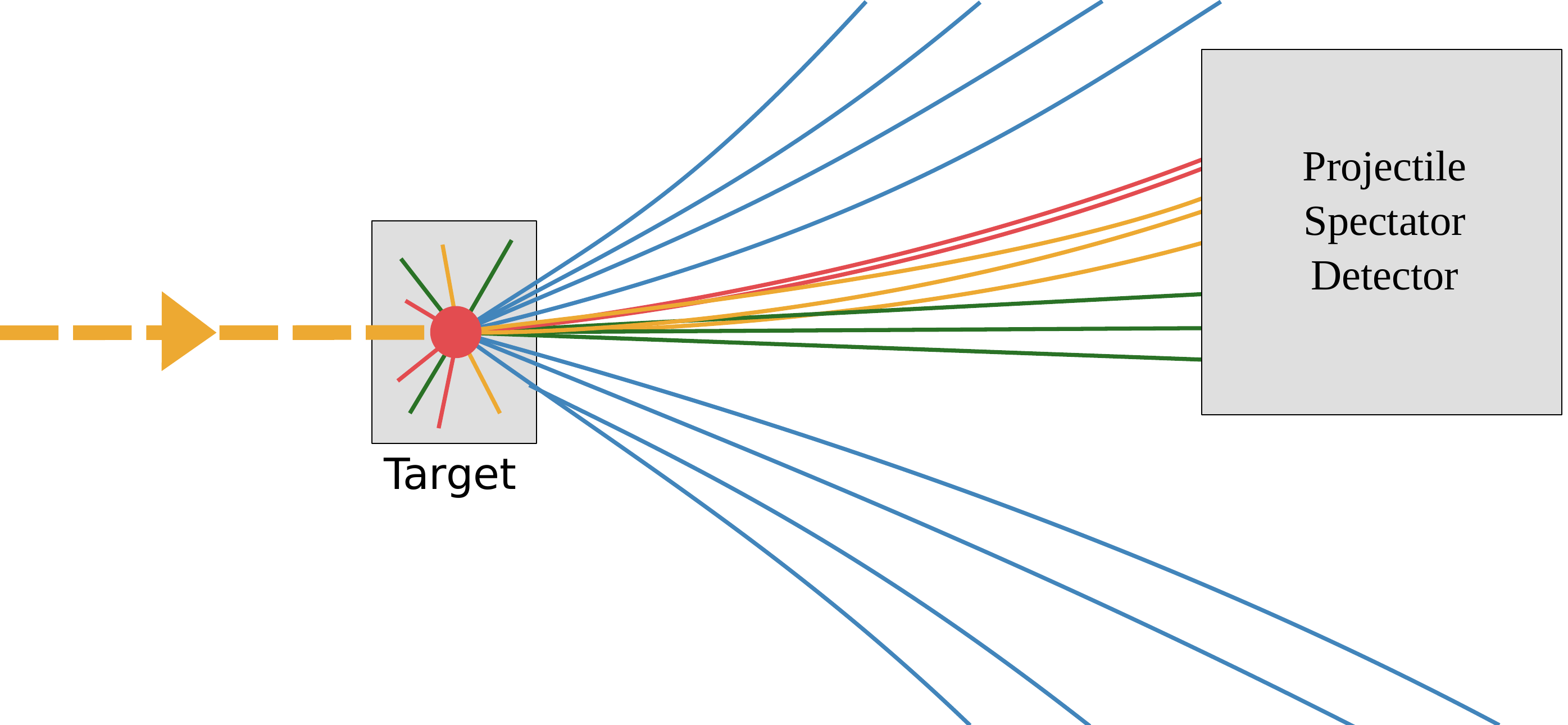}
	\caption[]
	{
		The sketch of a nucleus-nucleus collision as seen by a fixed-target experiment.
		{The incoming beam particle (marked with thick dashed orange line \textcolor{orange}{\dashedLine}) interacts inelastically with a target nucleus. Projectile spectators: protons (\textcolor{red}{\solidLine}), neutrons (\textcolor{green}{\solidLine}) and fragments (\textcolor{orange}{\solidLine}) propagate to the forward calorimeter. Newly produced hadrons' trajectories (\textcolor{blue}{\solidLine}) are bent and hadrons propagate to tracking detectors.}
	}
	\label{fig:FixedTarget}
\end{figure}

Typically, fixed-target experiments - like NA49 and \NASixtyOne at the CERN SPS - cover mostly the forward hemisphere in the 
center-of-mass system.
An advantage of the fixed-target geometry is that it allows to
select  collisions using the measured energy of spectators from the beam nucleus independently from measurements of the produced particles, see Fig.~\ref{fig:FixedTarget} for illustration.
This selection is referred to as \emph{centrality} selection.
It is important to note that the measurement of target spectators is usually impossible as 
most of them are fully stopped inside the target material.

\begin{figure}[ht]
	\centering
	\includegraphics[width=0.60\textwidth]{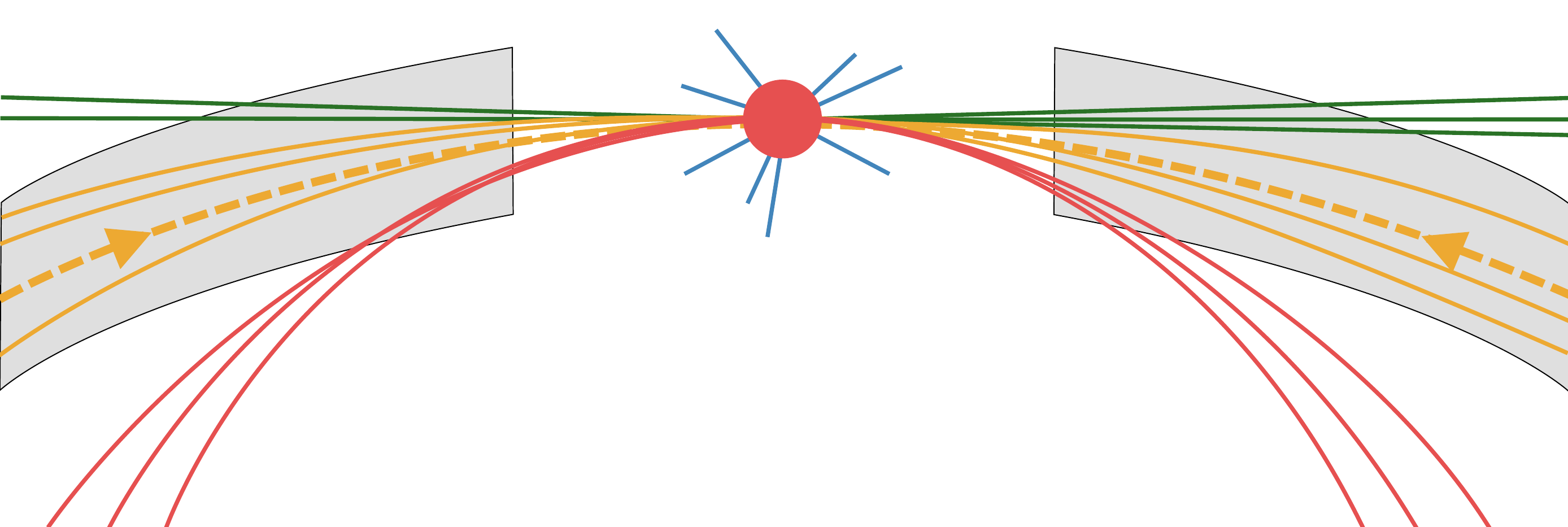}
	\caption[]
	{
		{The sketch of a nucleus-nucleus collision as seen by collider experiments. The incoming beam particles (marked with thick dashed orange line \textcolor{orange}{\dashedLine}) interact inelastically with each other.
		Measurements of \emph{left} and \emph{right} spectators are possible under the same experimental conditions. However, only free nucleon spectators (\textcolor{green}{\solidLine} and \textcolor{red}{\solidLine}) -- in central collisions about 50\% of all nucleons~\cite{Appelshauser:1998tt} -- can be measured. Fragments (\textcolor{orange}{\solidLine}) follow approximately the beam trajectories and they are difficult to measure. Newly produced hadrons (\textcolor{blue}{\solidLine}) propagate to other detectors.}
	}
	\label{fig:Collider} 
\end{figure}

On the other hand, a typical 
collider experiment -- like STAR at BNL RHIC and ALICE at CERN LHC -- has practically energy-independent rapidity acceptance, but without the low transverse momentum region. The track density in the detector increases only moderately with the
collision energy. However, \emph{left} and \emph{right} spectator regions are only partly accessible to measurements and the collision selection is usually based on the multiplicity of produced particles, see Fig.~\ref{fig:Collider} for illustration. Thus, quantities used to select events and study the properties of particle production are correlated by the physics of particle production.~This fact complicates the interpretation of the results.

\clearpage

\section{Two methods to remove \emph{volume} fluctuations}
\label{sec:MethodsTests}

In this section, the following
two popular methods to reduce the impact of the \emph{volume} fluctuations -- the unwanted component in Eq.~\ref{eq:wnm:varN}, are discussed:
\begin{enumerate}[(i)]
    \item selection of the most \emph{central} collisions,
    \item use of strongly intensive quantities.
\end{enumerate}


\subsection{The selection of the most \emph{central} collisions}
\label{sec:selection}

\begin{figure}[ht]
	\centering
	\includegraphics[width=0.80\textwidth]{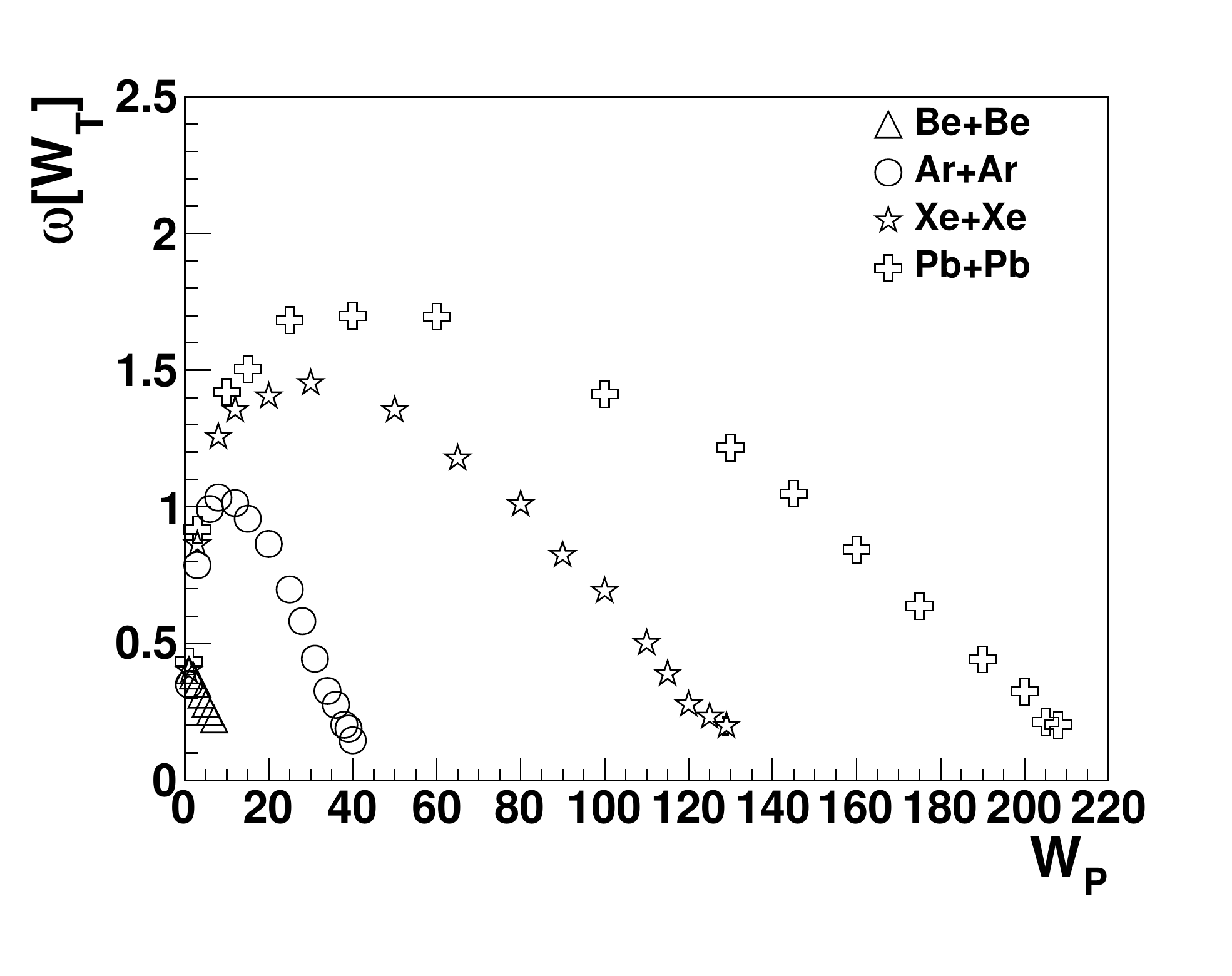}
	\caption[]
	{
	 Scaled variance of the distribution of the number  of target wounded nucleons $W_T$ 
	 as a function of the number of projectile wounded 
	nucleons $W_P$ calculated within the HIJING~\cite{Wang:1991hta}
    implementation of the Wounded Nucleon Model~\cite{Bialas:1976ed}. 
    The results for
    $^{7}Be+{}^{7}Be$,  $^{40}Ar+{}^{40}Ar$,  $^{129}Xe+{}^{129}Xe$ and  $^{208}Pb+{}^{208}Pb$ collisions at 19\AGeVc are presented.
    
    }
	\label{fig:omega_W}
\end{figure}

To limit the unwanted component in fixed-target experiments, collisions with the smallest number of projectile spectators are selected. 
This is done with collision-by-collision measurement of a forward energy that is 
predominantly the energy of projectile spectators, see Fig.~\ref{fig:FixedTarget}.
In order to simplify, let us assume that only collisions with zero number of projectile
spectators were selected and thus $W_P = A_P$, where $A_P$ is the nuclear mass number
of projectile nucleus.
Then, it appears that for collisions of sufficiently large 
nuclei of similar nuclear mass number, the number of target wounded nucleons is also fixed. 
This is demonstrated
in Fig.~\ref{fig:omega_W} where results obtained within the HIJING~\cite{Wang:1991hta}
implementation of the Wounded Nucleon Model~\cite{Bialas:1976ed} are shown.
These results agree with the predictions of the HSD and UrQMD models~\cite{Konchakovski:2005hq}.
Thus, the total number of wounded nucleons $W = W_P + W_T$ is approximately fixed for very 
\emph{central} collisions and its scaled variance is close to zero so
the unwanted component in Eq.~\ref{eq:wnm:varN} is eliminated.

\clearpage

\subsection{The use of strongly intensive quantities} 
\label{sec:siq}

Since even for the most \emph{central} collisions the \emph{volume} fluctuations 
cannot be fully eliminated, it is important
to minimise their effect further by defining suitable fluctuation measures.
It appears that for the WNM and the SM(IB-GCE) models~\cite{Gazdzicki:1992ri,Gorenstein:2011vq,Sangaline:2015bma} fluctuation measures
independent of the \emph{volume} fluctuations can be constructed using
moments of the distribution of two extensive quantities.

As the simplest example, let us consider multiplicities
of two different types of hadrons, $A$ and $B$.
Their mean multiplicities are proportional to $W$:
\begin{equation}
\label{eq:wnm:mean}
 \langle A \rangle \sim W~, ~~~~~ \langle B \rangle \sim W. 
\end{equation}
Obviously the ratio of mean multiplicities is independent of  $W$.
Moreover, the ratio $\langle A \rangle \ / \langle B \rangle $ is
independent of $P(W)$, where $P(W)$ is
the probability distribution of $W$ for a selected set of collisions. 
The quantities which have the latter property are called strongly
intensive quantities~\cite{Gorenstein:2011vq}. Such quantities are useful in experimental studies of fluctuations in A+A collisions as they eliminate the influence of 
a usually poorly known distribution of $W$.

More generally, $A$ and $B$ can be any extensive
event quantities such as the sum of transverse momenta, the net charge or
the multiplicity of particles of a given type.
The scaled variance of $A$ and $B$ and the mixed second
moment $\langle AB \rangle$ calculated within 
the WNM~\cite{Gorenstein:2011vq} read:
\begin{equation}
 \omega[A] = \omega^*[A] + \langle A \rangle / \langle W \rangle \cdot\omega[W] ~, \\ 
\label{eq:wnm:varA}
\end{equation}
\begin{equation}
 \omega[B] = \omega^*[B] + \langle B \rangle / \langle W \rangle \cdot\omega[W] ~, \\
\label{eq:wnm:varB}
\end{equation}
\begin{equation}
 \langle AB \rangle = \langle AB \rangle^ * \langle W \rangle +
 \langle A \rangle  \langle B \rangle \langle W \rangle^2 \cdot
 (\langle W^2 \rangle - \langle W \rangle )~,
\label{eq:wnm:AB}
\end{equation}
where the quantities denoted by $^*$
are quantities calculated at a fixed \emph{volume}.

From Eqs.~\ref{eq:wnm:varA}-\ref{eq:wnm:AB} 
it follows~\cite{Gorenstein:2011vq,Gazdzicki:2013ana} that
\begin{equation}\label{eq:delta}
 \Delta[A,B]
 ~=~ \frac{1}{C_{\Delta}} \Big[ \langle B\rangle\,
      \omega[A] ~-~\langle A\rangle\, \omega[B] \Big]
\end{equation}
and
\begin{equation}\label{eq:sigma}
  \Sigma[A,B]
 ~=~ \frac{1}{C_{\Sigma}}\Big[
      \langle B\rangle\,\omega[A] ~+~\langle A\rangle\, \omega[B] ~-~2\left(
      \langle AB \rangle -\langle A\rangle\langle
      B\rangle\right)\Big]
\end{equation}
are independent of $P(W)$ in the WNM.
Here, the normalisation factors
$C_{\Delta}$ and $C_{\Sigma}$ are required to be proportional to the first moments
of any extensive quantity.
In Ref.~\cite{Gazdzicki:2013ana} a specific choice of the $C_{\Delta}$ and
$C_{\Sigma}$ normalisation factors was proposed which makes the quantities
$\Delta[A,B]$ and $\Sigma[A,B]$ dimensionless and  leads to
$\Delta[A,B] = \Sigma[A,B] = 1$ in the independent particle
model (IPM)~\cite{Gazdzicki:2013ana}. 
This normalisation is referred to as the IPM normalisation and it is used here.

Thus, $\Delta[A,B]$ and $\Sigma[A,B]$ are strongly intensive quantities
which measure fluctuations of $A$ and $B$, 
i.e. they are sensitive to second moments of the distributions
of the quantities $A$ and $B$. 
The results on $\Delta[A,B]$ and $\Sigma[A,B]$ are referred to as
the results on $A - B$ fluctuations, e.g., transverse momentum - multiplicity fluctuations.
The analogous quantities called strongly intensive cumulants allow to measure fluctuations of higher order moments~\cite{Sangaline:2015bma}. The first four are defined as:
\begin{eqnarray}
    \kappa_{1}^{*}[A,B] & = & \frac{\langle A \rangle}{\langle B \rangle}\nonumber \\
\kappa_{2}^{*}[A,B] & = & \frac{\langle A^{2} \rangle}{\langle B \rangle}-\frac{\langle A \rangle\langle AB \rangle}{\langle B \rangle^{2}}\nonumber \\
\kappa_{3}^{*}[A,B] & = & \frac{\langle A^{3} \rangle}{\langle B \rangle}-\frac{2\langle A^{2} \rangle\langle AB \rangle+\langle A \rangle\langle A^{2}B \rangle}{\langle B \rangle^{2}}+\frac{2\langle A \rangle\langle AB \rangle^{2}}{\langle B \rangle^{3}}\label{eq:first-four-strongly intensive cumulants}\\
\kappa_{4}^{*}[A,B] & = & \frac{\langle A^{4} \rangle}{\langle B \rangle}-\frac{3\langle A^{3} \rangle\langle AB \rangle+\langle A \rangle\langle A^{3}B \rangle}{\langle B^{2} \rangle}-\frac{3\langle A^{2} \rangle\langle A^{2}B \rangle}{\langle B^{2} \rangle}+\nonumber \\
 &  &  \frac{6\langle A^{2} \rangle\langle AB \rangle^{2}+6\langle A \rangle\langle A^{2}B \rangle\langle AB \rangle}{\langle B^{3} \rangle}-\frac{6\langle A \rangle\langle AB \rangle^{3}}{\langle B^{4} \rangle}\nonumber 
\end{eqnarray}

Because of their construction, strongly intensive measures of fluctuations
require two extensive quantities. This, in general, hampers
a straight-forward interpretation of the experimental results. However, under
certain conditions the $\Delta$ quantity can be used to
obtain the scaled variance of the extensive quantity $A$ separately.  

Let $A$ be an extensive quantity, e.g., selected for its sensitivity
to critical fluctuations. Then choose a quantity
$B$ such that  
$ B  \sim W$ and denote it as $B_W$.
It is easy to show~\cite{Gazdzicki:2015ska} that the strongly intensive measures 
$\Delta_{B}[A,B]$ and $\Sigma_{B}[A,B]$
(equal to $\Delta[A,B]$ and $\Sigma_{B}[A,B]$ with the normalisation 
$C_{\Delta} = \langle B \rangle \sim \langle W \rangle$)
obey the relation:
\begin{equation}
    \Delta_{B}[A,B] = \Sigma_{B}[A,B] =  \omega^*[A] ~.
 \label{eq:delta_c}
\end{equation}
Thus, $\Delta_{B}[A,B]$ is equal to the scaled variance $\omega[A]$ 
for a fixed number of wounded nucleons (see Eq.~\ref{eq:wnm:varA}).
Similar relations can be found for strongly intensive cumulants of any order:
\begin{equation}
	\frac{\kappa_{n}^{*}}{\kappa_{k}^{*}}[A,B_{W}] = \frac{\kappa_{n}[A]}{\kappa_{k}[A]}~.
	\label{eq:sic}
\end{equation}
In the derivation of Eqs.~\ref{eq:delta_c} and~\ref{eq:sic} one assumes 
the validity of Eq.~\ref{eq:wnm:varA}
which needs to be investigated case-by-case.

\section{Numerical tests}
\label{sec:tests}

Numerical tests of 
the methods to eliminate the \emph{volume} fluctuations introduced above are presented in this
section. The simulations were performed using the HIJING~\cite{Wang:1991hta} implementation of the Wounded Nucleon Model:
\begin{enumerate}[(i)]
\item
$^{40}$Ar+$^{40}$Ar collisions at 150\AGeVc were generated. 
This reaction closely corresponds to 
data recorded by \NASixtyOne at the CERN SPS~\cite{Mackowiak-Pawlowska:2020glz}.
\item
for each collision, number of projectile and target wounded nucleons and impact
parameter $b$ are stored.
\item
number of particles produced by a given wounded nucleon $N$ is drawn from the binomial distribution with $N_{max}=2$ and $p=0.5$.
Moments of this distribution are:
$\avg{N} = 1$,  $\omega[N] = 0.5$,  $\kappa_{3}[N]/\kappa_{2}[N] = 0$ and  $\kappa_{4}[N]/\kappa_{2}[N] = -0.5$.

\end{enumerate}

\subsection{Selecting the most \emph{central} collisions}
\label{sec:test:violent}

\begin{figure}[ht]
	\centering
	\includegraphics[width=0.45\textwidth]{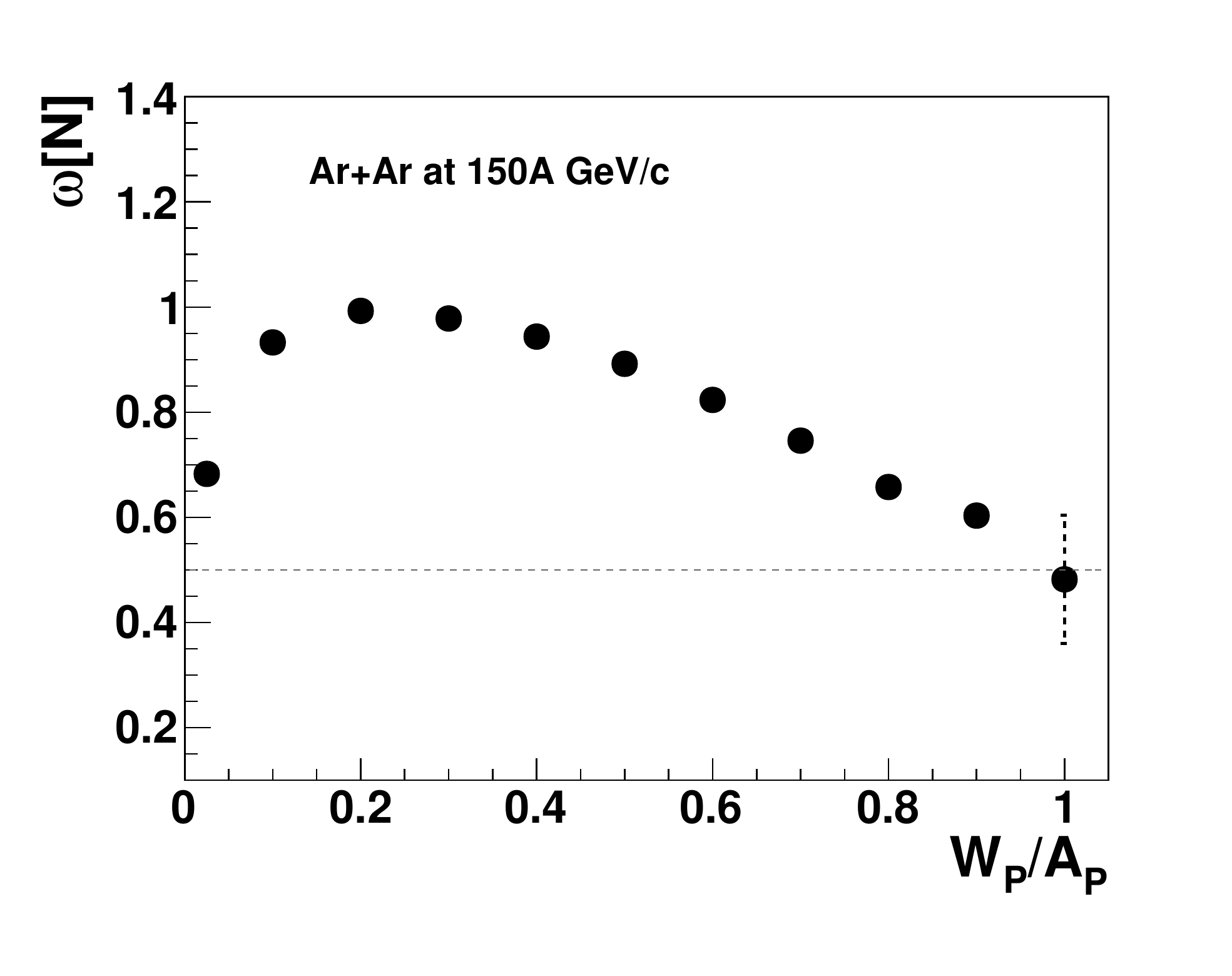}	\includegraphics[width=0.45\textwidth]{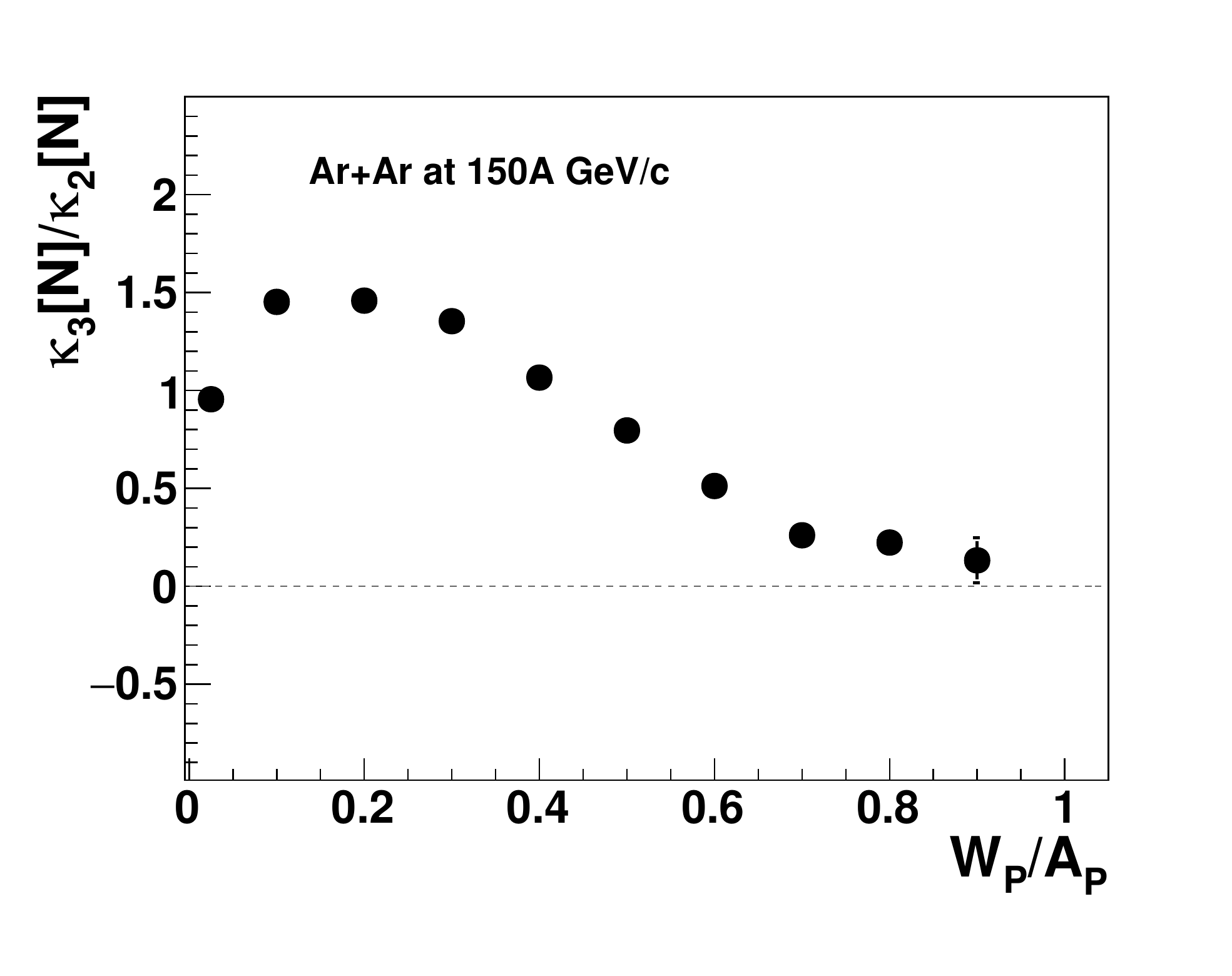}	\includegraphics[width=0.45\textwidth]{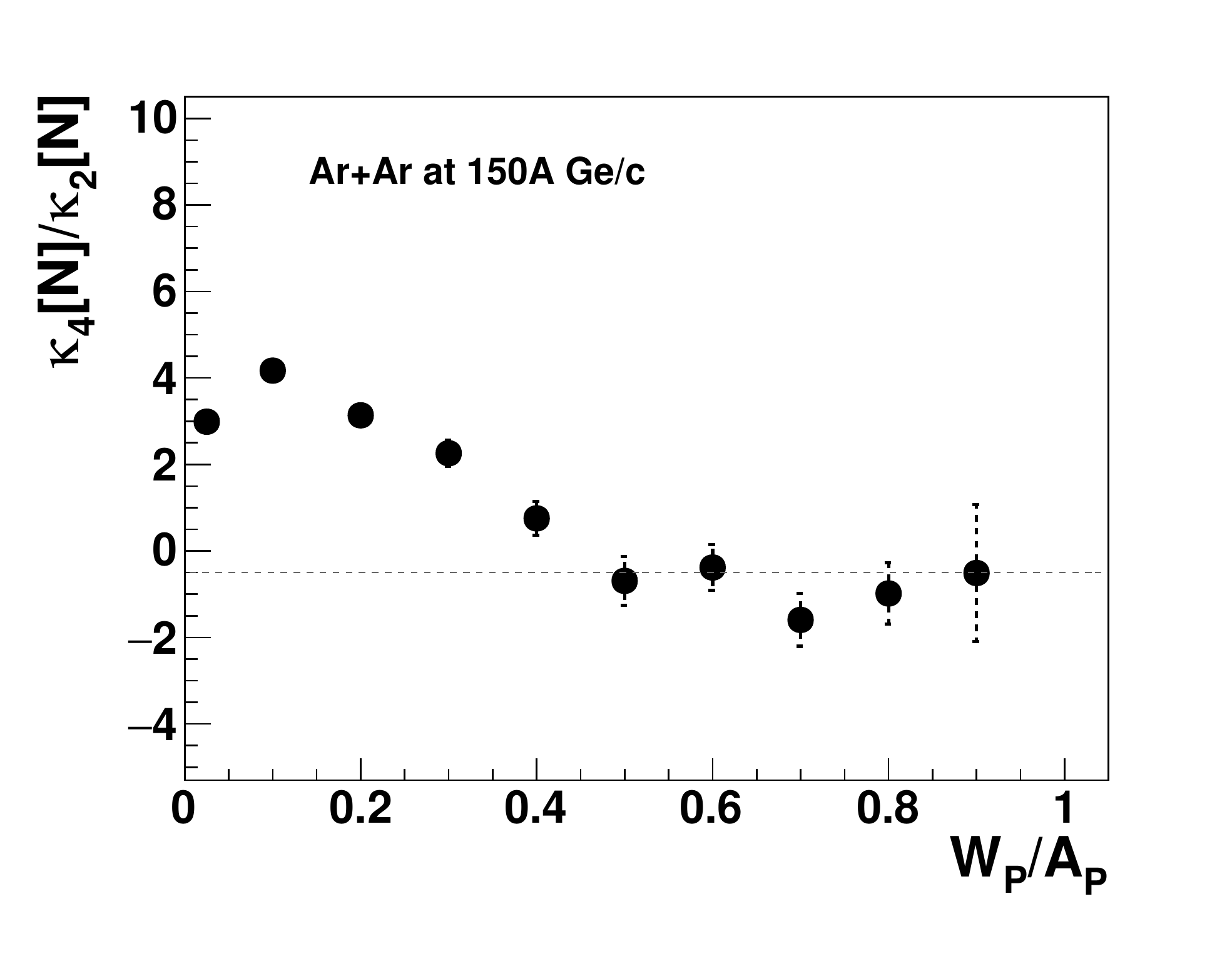}
	\caption[]
	{
	    The dependence of $\omega[N]$, $\kappa_3[N]/\kappa_2[N]$ and $\kappa_4[N]/\kappa_2[N]$ 
	    on the ratio $W_P/A_P$ within the Wounded Nucleon Model with input defined in Sec.~\ref{sec:tests}. The reference values for any fixed number of wounded nucleons $W = const$ are shown by dashed lines. The calculations were performed for $^{40}$Ar+$^{40}$Ar collisions at 150\AGeVc. 
	}
	\label{fig:TestViolent}
\end{figure}

Figure~\ref{fig:TestViolent} shows the dependence of $\omega[N]$,
$\kappa_3[N]/\kappa_2[N]$ and $\kappa_4[N]/\kappa_2[N]$ on the ratio $W_P/A_P$.
The quantities approach the corresponding value for a fixed number of wounded nucleons with
$W_P/A_P \to 0$.
It is important to note that  only $\approx 0.0007\%$ of all inelastic collisions have $W_P = A_P$.

It can be concluded that the selection of collisions with $W_P = A_P$ significantly reduces the effect of the \emph{volume} fluctuations, however it is  at the cost of  reduction of event statistics.
The remaining  bias can be corrected for using a model-dependent correction.
The uncertainty of this correction will contribute to the systematic uncertainty of the final results.

\subsection{Using strongly intensive quantities}
\label{sec:test:siq}

\begin{figure}[ht]
	\centering
    		\includegraphics[width=0.45\textwidth]{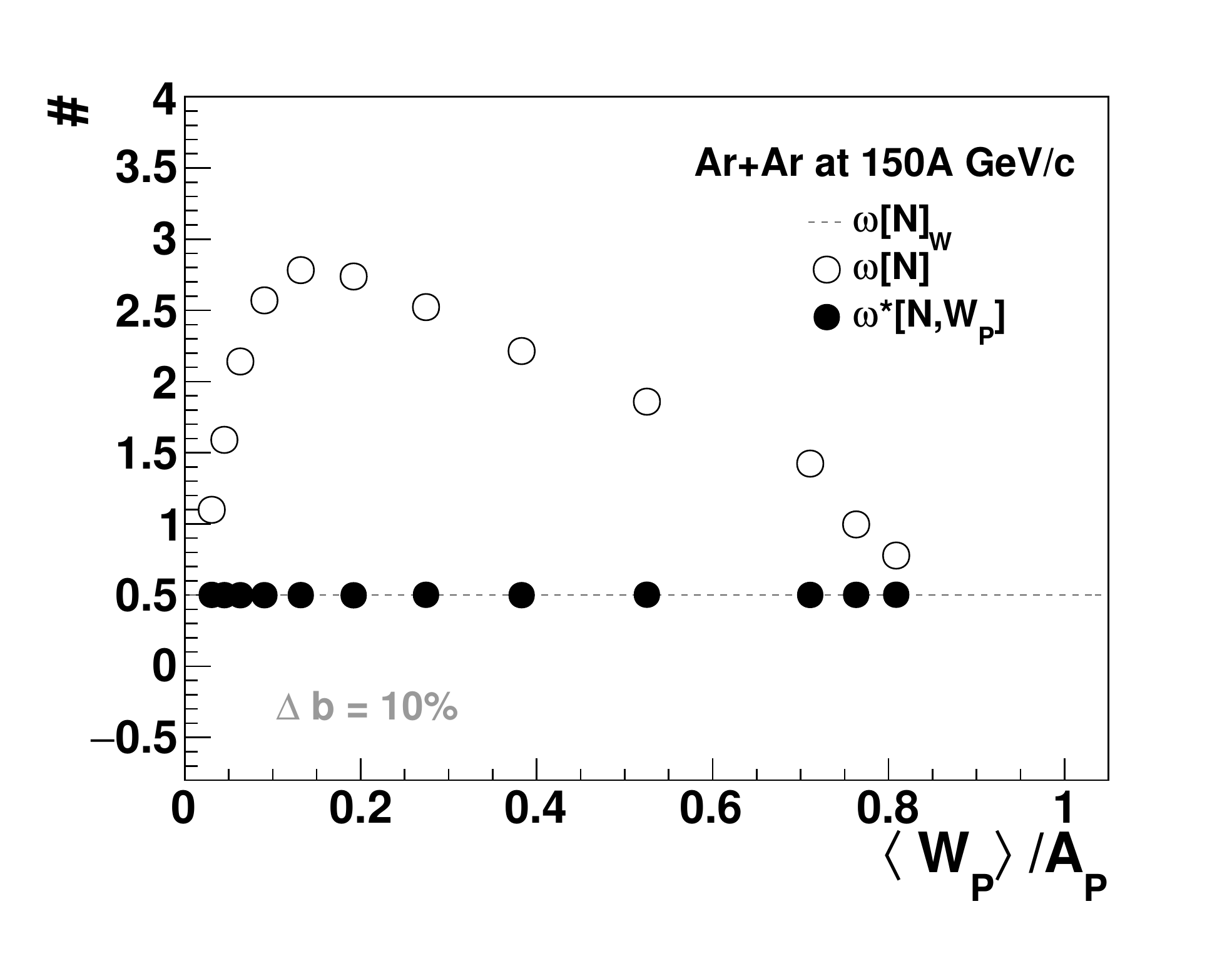}	\includegraphics[width=0.45\textwidth]{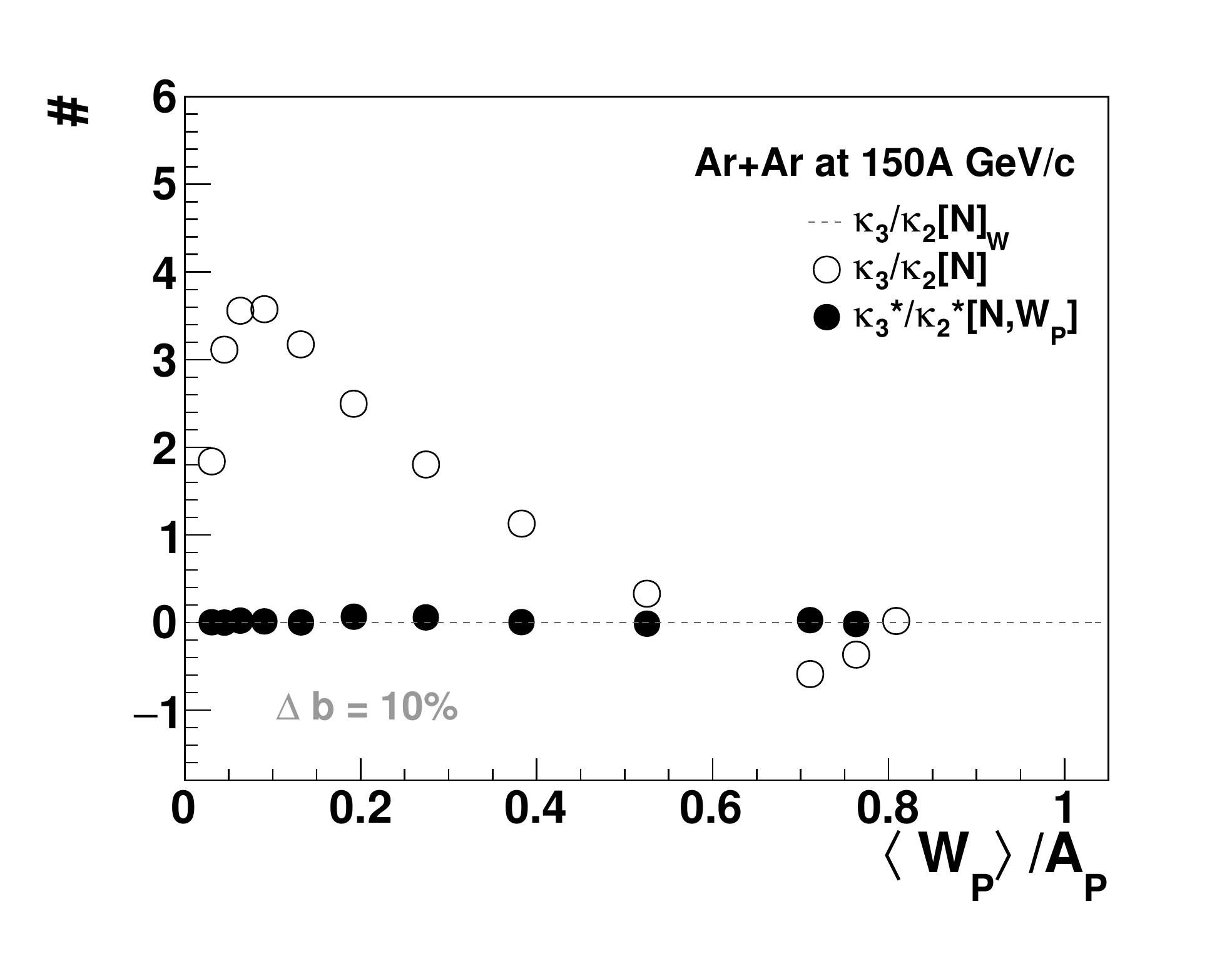}	\includegraphics[width=0.45\textwidth]{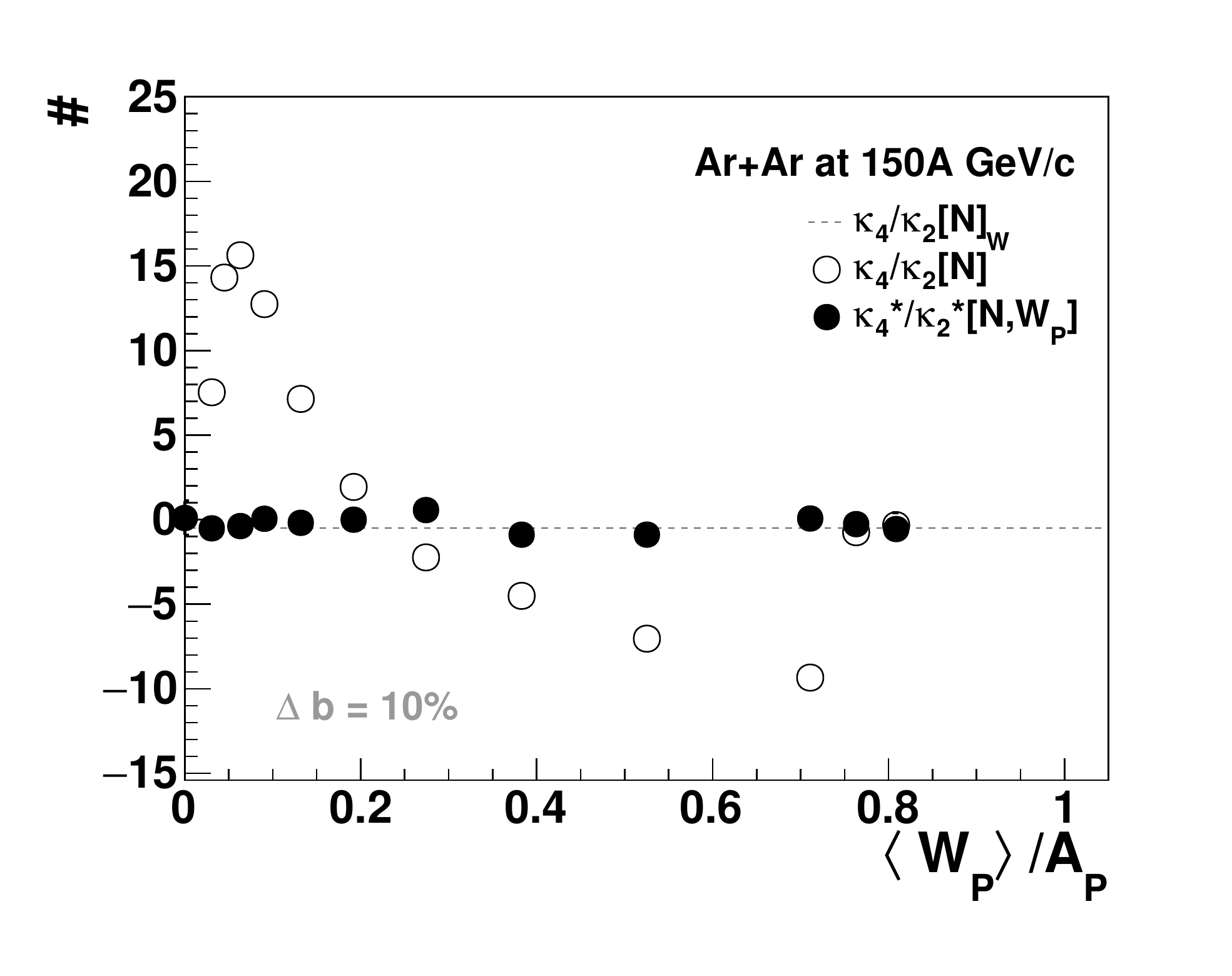}
	\caption[]
	{
	    The dependence of $\omega^*[N,W_P]$, $\kappa_3^*[N,W_P]/\kappa_2^*[N,W_P]$ and $\kappa_4^*[N,W_P]/\kappa_2^*[N,W_P]$ (full circles) as well as $\omega[N]$, $\kappa_3[N]/\kappa_2[N]$ and $\kappa_4[N]/\kappa_2[N]$ (open circles) on the ratio 
	    $\avg{W_P}/A_P$ within the Wounded Nucleon Model with input defined in Sec.~\ref{sec:tests}. Results are obtained in bins of $b$. The most right two points which correspond to $\Delta b$ equal to $5\%$ and $1\%$. The values for $W = const$ are shown by dashed lines. 
	}
	\label{fig:TestSIQ3}
\end{figure}

Strongly intensive quantities were proposed with the aim to reduce the intrinsic limitation of the  method
based on the selection of \emph{central} events which may lead to significant systematic and statistical uncertainties.
Figure~\ref{fig:TestSIQ3} shows the dependence of intensive and strongly intensive quantities on the ratio of $\langle W_P \rangle/A_P$. Here, collisions were selected using collision impact parameter. As expected, strongly intensive quantities are equal or are close to the corresponding values for fixed $W$. Unlike strongly intensive quantities their intensive partners also shown in Fig.~\ref{fig:TestSIQ3} significantly depend on the impact parameter selection.
Thus, it can be concluded that strongly intensive quantities together with the impact parameter
selection of collisions fully eliminates the effect of \emph{volume} fluctuations.
Unfortunately, this is not the solution of the problem.
The collision impact parameter is not a measurable quantity.
So, calculating strongly intensive quantities for all inelastic collisions should be considered.
Within the WNM and SM(IB-GCE) models, strongly intensive quantities for those collisions are equal to the corresponding quantities for fixed $W$.
However, in general, the models are not valid in the full range of the impact parameter.

\begin{figure}[ht]
	\centering
    		\includegraphics[width=0.45\textwidth]{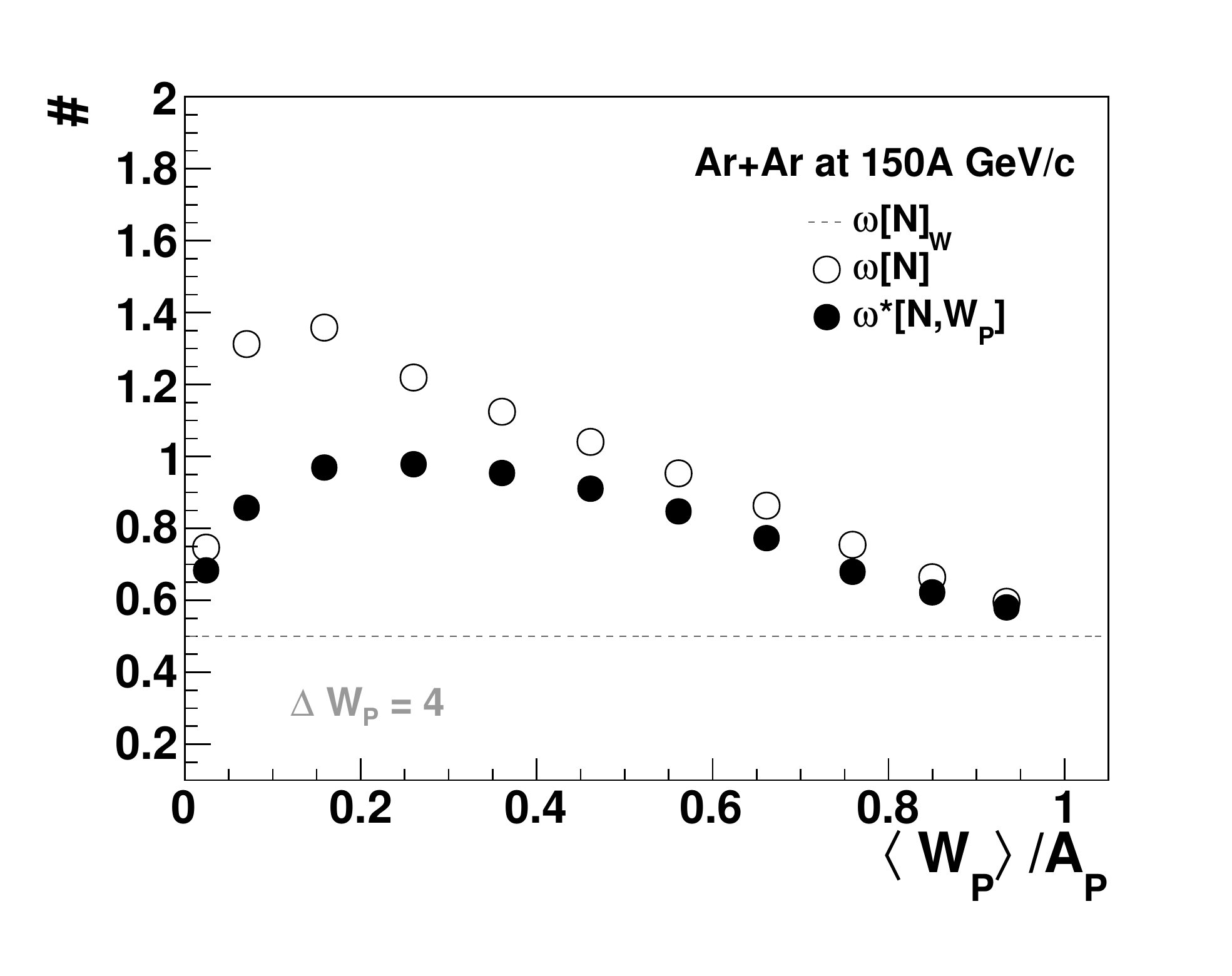}	\includegraphics[width=0.45\textwidth]{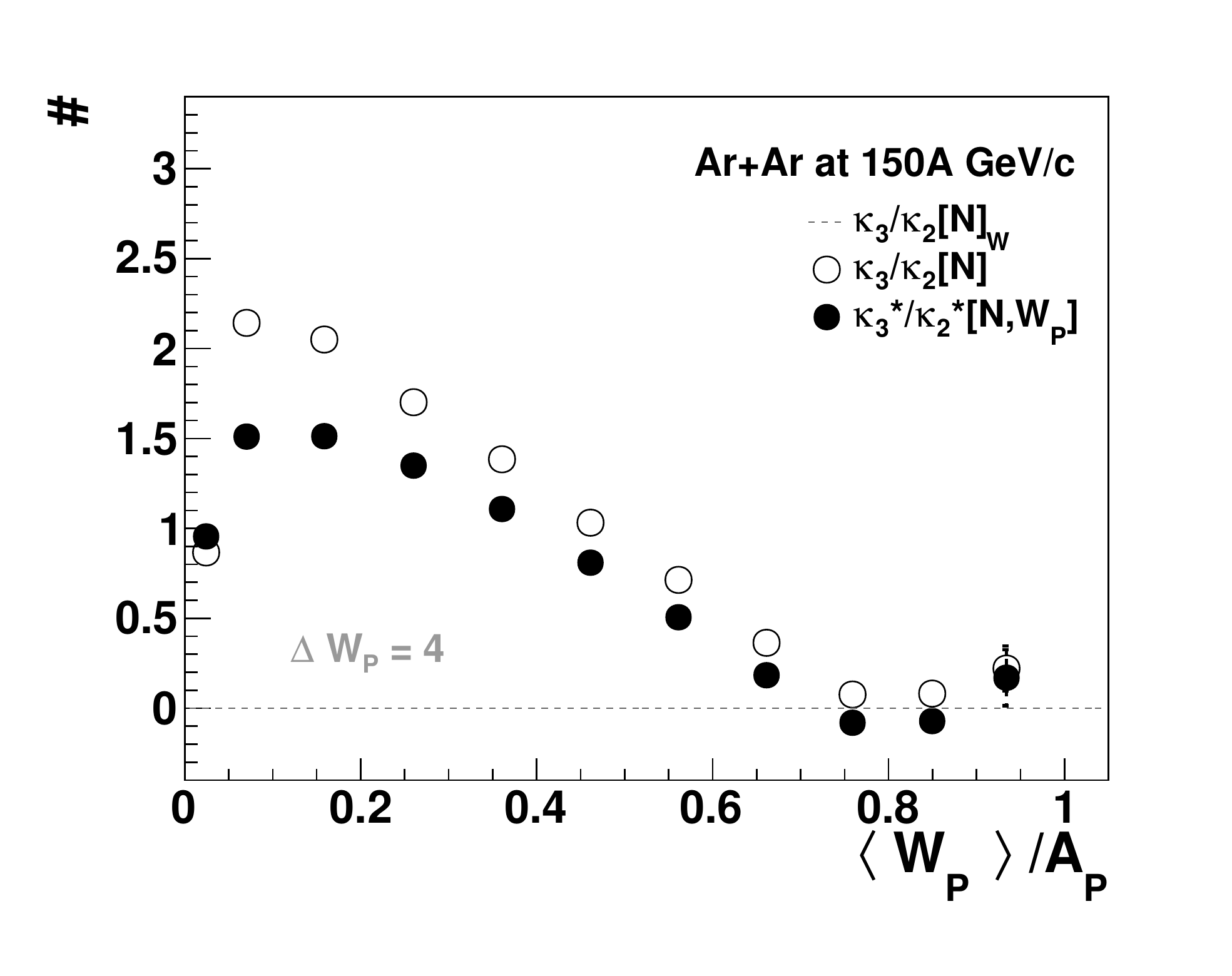}	\includegraphics[width=0.45\textwidth]{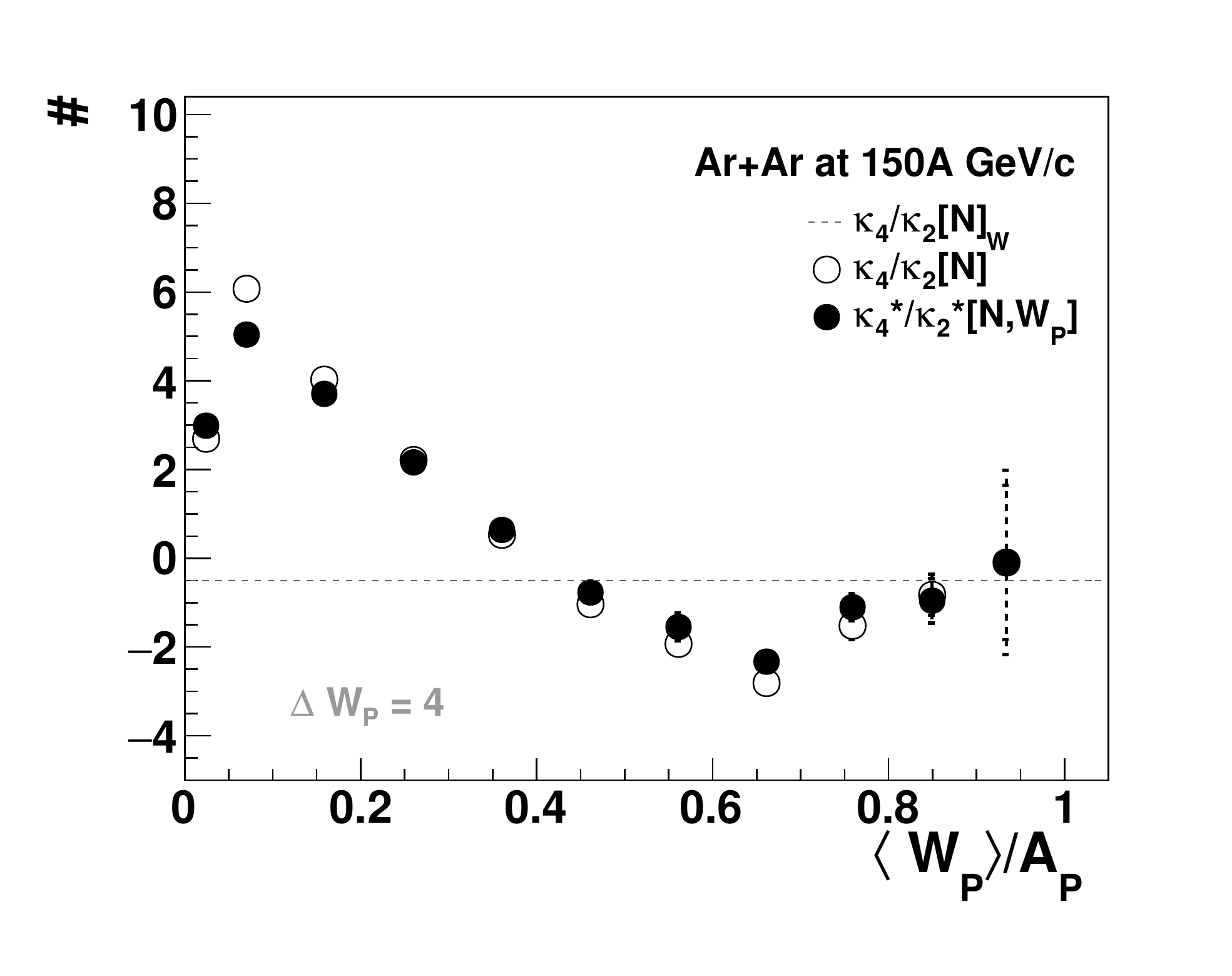}
	\caption[]
	{
	    The dependence of $\omega^*[N,W_P]$, $\kappa_3^*[N,W_P]/\kappa_2^*[N,W_P]$ and $\kappa_4^*[N,W_P]/\kappa_2^*[N,W_P]$ (full circles) as well as $\omega[N]$, $\kappa_3[N]/\kappa_2[N]$ and $\kappa_4[N]/\kappa_2[N]$ (open circles) on the ratio 
	    $\avg{W_P}/A_P$  within the Wounded Nucleon Model with input defined in Sec.~\ref{sec:tests}. Results are obtained in bins of $W_P$. The values for  $W = const$ are shown by dashed lines. 
	   
	}
	\label{fig:TestSIQ2}
\end{figure}

Consequently, the method of the event selection based on the number of projectile 
wounded nucleons needs to be used.
The results calculated in bins of $W_{P}$ are shown in Fig.~\ref{fig:TestSIQ2}. 
In this case strongly intensive quantities, in general, also deviate from the corresponding values for fixed $W$. They approach them only for the most \emph{central} collisions,
$W_P/A_P \to 1$.
This is due to the introduced correlation when events are selected on the same quantity used to calculate strongly intensive quantities.
This can be solved by defining strongly intensive quantities using two extensive quantities related to particle production properties.
Particle multiplicity and 
transverse momentum~\cite{Anticic:2015fla, Gorenstein:2013nea} are the most popular examples of these quantities.
However, when the goal is to obtain moments of multiplicity distribution for fixed $W$
there is no significant advantage of using strongly intensive quantities.
Similarly, intensive quantities have to be calculated in the most \emph{central} collisions
to approach the unbiased results.


\clearpage

\section{Summary}
\label{sec:summary}

The paper addresses a currently important question of measuring event-by-event 
particle number fluctuations in nucleus-nucleus collisions unbiased by fluctuations of the collision \emph{volume}.
Two methods to remove the influence of the \emph{volume} fluctuations in fixed target experiments are reviewed and tested.
Some of the limitations of \emph{strongly intensive quantities} in these types of analysis are shown.
The results indicate the need to select the most \emph{central} collisions using the
number of projectile spectators.

\begin{acknowledgments} The authors thank K.~Grebieszkow and A.~Seryakov as well as other participants of Ion and PSD meetings of the \NASixtyOne collaboration. This work was supported by the Polish National Science Centre grants 2016/21/D/ST2/01983, 2018/30/A/ST2/00226 and 2019/32/T/ST2/00432 as well as the German Research Foundation grant GA1480\slash 8-1. MMP studies were also funded by IDUB-POB-FWEiTE-1 project granted by
Warsaw University of Technology under the program Excellence Initiative: Research University
(ID-UB).
\end{acknowledgments}

\bibliographystyle{ieeetr}
\bibliography{references}

\end{document}